\documentclass[conference]{IEEEtran}
\IEEEoverridecommandlockouts
\usepackage{cite}
\usepackage{amsmath,amssymb,amsfonts}
\usepackage{algorithmic}
\usepackage{graphicx}
\usepackage{textcomp}
\usepackage{xcolor}
\usepackage{tabularx}
\usepackage{makecell}
\usepackage{array}
\usepackage{float}
\usepackage{caption}
\usepackage{subcaption}
\usepackage{multicol}
\usepackage{algorithm}
\usepackage{booktabs}
\usepackage[a4paper, top=1.9cm, bottom=4.3cm, left=1.3cm, right=1.3cm]{geometry}
\def\BibTeX{{\rm B\kern-.05em{\sc i\kern-.025em b}\kern-.08em
    T\kern-.1667em\lower.7ex\hbox{E}\kern-.125emX}}

\title{FilterLoss: A Transfer Learning Approach for Communication Scene Recognition}

\author{
	Jiasong~Han\textsuperscript{1},~Yufei~Feng\textsuperscript{1},~Xiaofeng~Zhong\textsuperscript{1,2,3,*}\\
	\IEEEauthorblockA{
		\textsuperscript{1}Department of Electronic Engineering, Tsinghua University, Beijing 100084, China\\
		\textsuperscript{2}Beijing National Research Center for Information Science and Technology, Beijing 100084, China\\
        \textsuperscript{3}State Key laboratory of Space Network and Communications (Tsinghua University), Beijing 100084, China\\
        \textsuperscript{}Email: \{hanjs22, fengyf23\}@mails.tsinghua.edu.cn, zhongxf@tsinghua.edu.cn\\
        \textsuperscript{*}Corresponding author: Xiaofeng Zhong}
}        

\begin{document}

\maketitle

\begin{abstract}
Communication scene recognition has been widely applied in practice, but using deep learning to address this problem faces challenges such as insufficient data and imbalanced data distribution. To address this, we designed a weighted loss function structure, named FilterLoss, which assigns different loss function weights to different sample points. This allows the deep learning model to focus primarily on high-value samples while appropriately accounting for noisy, boundary-level data points. Additionally, we developed a matching weight filtering algorithm that evaluates the quality of sample points in the input dataset and assigns different weight values to samples based on their quality. By applying this method, when using transfer learning on a highly imbalanced new dataset, the accuracy of the transferred model was restored to 92.34\% of the original model’s performance. Our experiments also revealed that using this loss function structure allowed the model to maintain good stability despite insufficient and imbalanced data.

\end{abstract}


\section{INTRODUCTION}
In the communication scene recognition project, smartphones' Micro-Electro-Mechanical System (MEMS) sensors, including accelerometers, gyroscopes, and magnetometers, are used to identify users' communication environments. By analyzing real-time data, the system can determine user mobility, environmental context (e.g., indoor, outdoor, in a vehicle), and activities (e.g., walking, running, public transportation). These insights are crucial for optimizing communication systems, as different scenes significantly affect signal propagation.

Mobile devices' rapid development challenges current communication scene recognition algorithms. Single-source data limits cross-sensor adaptability. As TABLE \ref{tab:decline} shows, testing an SHL-2018-trained algorithm \cite{a0} on Beijing-2024 data \cite{b10} reveals significant accuracy loss. Expanding training data dimensionality helps but remains insufficient.

\begin{table}[ht]
\centering
\begin{tabular}{|m{1.4cm}|m{1.8cm}|m{1.8cm}|}
\hline
Dimension & Train \& Test on SHL-2018 & Train on SHL-2018 \& Test on Beijing-2024 \\
\hline
\centering 8 & \multicolumn{1}{|c|}{58.00\%} & \multicolumn{1}{|c|}{12.39\%} \\
\hline
\centering 16 & \multicolumn{1}{|c|}{61.67\%} & \multicolumn{1}{|c|}{15.83\%} \\
\hline
\centering 32 & \multicolumn{1}{|c|}{67.62\%} & \multicolumn{1}{|c|}{20.64\%} \\
\hline
\centering 64 & \multicolumn{1}{|c|}{63.53\%} & \multicolumn{1}{|c|}{26.83\%} \\
\hline
\end{tabular}
\caption{A trained model's accuracy significantly declines when the sampling equipment undergoes generational updates}
\label{tab:decline}
\end{table}

Most communication scene recognition algorithms \cite{a1,a2} use the SHL-2018 dataset, which provides labeled MEMS sensor data from smartphones. However, evolving sensor technology and variations across smartphone models introduce significant data and noise discrepancies \cite{b1,b2,b3}, undermining algorithm generalization. TABLE \ref{tab:Android} shows notable sensor performance differences across device types and years, making cross-device adaptation a key challenge.

\begin{table}[ht]
\centering
\begin{tabular}{|c|c|c|c|}
\hline
Accelerometers & Release Date & Digital Resolution & Noise Density \\
\hline
BMA253\cite{b1} & October 2021 & 12bit & 220µg/$\sqrt{Hz}$ \\
\hline
BMA580\cite{b2} & May 2024 & 16bit & 120µg/$\sqrt{Hz}$ \\
\hline
\multicolumn{4}{c}{ } \\
\hline
Magnetometer & Release Date & Digital Resolution & Zero-B Offset\\
\hline
BMM150\cite{b3} & April 2020 & 14bit & ±40µT \\
\hline
BMM350\cite{b3} & March 2024 & 16bit & ±25µT \\
\hline
\end{tabular}
\caption{Intergenerational improvement of sensor performance in android phones}
\label{tab:Android}
\end{table}

Training a new model for each smartphone is impractical due to high resource costs and data collection challenges. Instead, we propose transfer learning \cite{b14}, leveraging pretraining and fine-tuning to achieve generalization with limited datasets.

Another key issue is dataset imbalance, as walking/driving data far outweighs train travel. While undersampling and oversampling are common solutions \cite{b12,b13}, blindly adjusting samples risks amplifying noise or discarding critical data, ultimately harming accuracy \cite{b6}.

Therefore, in this paper, we propose a new loss function structure named FilterLoss incorporating a weight filter to effectively address issues related to small sample sizes and imbalanced dataset distributions. We pretrain a model on an open source dataset collected in 2018. This pretrained model, though performing good on 2018 sensor data, could not effectively work on lately-gathered sensor data. Therefore, we applied FilterLoss and fine-tuned the model using the Beijing-2024 dataset, achieving significant improvements. Specifically, we pass it through a weight filter that assigns different loss function weights to each sample. During fine-tuning, we update the last convolution layer's parameters using this loss function with different weights, achieving improved learning performance. Our experiments demonstrate that using the loss function with the weight filter effectively resolves challenges in communication scene recognition while maintaining low computational costs. Additionally, compared to traditional over sampling and under sampling strategies, our approach results in more stable metrics such as accuracy, F1 score, and loss function values.

In summary, our contributions are as follows:
\begin{itemize}
\item 
We propose a strategy named FilterLoss to address the issues of limited data and imbalanced distribution in transfer learning for communication scene recognition. By filtering sample points into different loss function weights, the accuracy of the transfer learning algorithm is restored to 92.34\% of its original level, and the results are more stable.
\item 
We also developed a corresponding weight assignment algorithm to determine the appropriate weight for each sample point. Compared to traditional binary methods for handling imbalanced sample distributions, this "soft" approach leads to more stable model performance and better boundary handling.
\item 
Our proposed method requires minimal data collection and computational costs. Training only takes three minutes on a 3090Ti GPU, and data collection can be performed using a smartphone, a tablet or a smart watch, making it applicable to most mobile communication devices.
\end{itemize}

\section{RELATED WORKS}

\subsection{Communication Scene Recognition}


The integration of deep learning with sensor technologies has significantly advanced communication scene recognition. He et al. \cite{b6} pioneered the application of ResNets for multi-sensor feature extraction, mitigating vanishing gradients and degradation issues. Krichen \cite{b6} emphasized the importance of data preprocessing, such as noise reduction and normalization, which informs our pipeline. Subsequent works have further refined the field by integrating contextual information \cite{b7} and hybrid CNN-RNN architectures \cite{b9} to capture spatiotemporal dependencies. For example, Gjoreski et al. \cite{b8} combined classical feature engineering with deep learning, while Feng et al. \cite{b10} achieved a notable 91.68\% accuracy using a ResNet-34-based method. Building on these contributions, our approach integrates residual learning, hybrid architectures, and advanced preprocessing for robust and accurate recognition.

Many studies focus on specific datasets, but adding sensors and data sources doesn't always improve research quality. The key issue often overlooked is generalizability—data from older models may not apply to newer ones. It's crucial to prioritize versatility and adaptability in research to maintain the practical value of findings across device generations.

\subsection{Transfer Learning}

Transfer learning (TL) effectively addresses data scarcity by transferring knowledge from a data-rich source domain to a target domain with limited labeled data. Pan and Yang \cite{b20} provide a foundational survey on TL, categorizing it into inductive, transductive, and unsupervised scenarios to highlight its versatility in adapting models to new tasks. Building on this, Weiss et al. \cite{b21} further explore techniques like domain adaptation, fine-tuning, and feature extraction, demonstrating how TL can enhance model generalization and reduce training time by leveraging related source domain information. These studies collectively underscore TL's importance in overcoming data limitations. Inspired by these contributions, our approach utilizes TL to improve the performance of communication scene recognition tasks under data-scarce conditions.

In wireless communications, TL has been instrumental in tasks such as network traffic prediction and channel state information (CSI) estimation. Chen et al. \cite{b22} examine the application of TL for wireless communications, demonstrating its potential in enhancing system performance. Zappone et al. \cite{b23} discuss the integration of deep learning and model-based approaches in wireless network design, emphasizing the role of TL in adapting models to new environments without extensive data collection. This body of work underscores TL's value in reducing the need for large target domain datasets, thereby lowering costs and accelerating deployment in practical applications.

These studies primarily address the issue of insufficient data, but in the context of communication scene recognition, the challenge is not only the lack of data but also the imbalance in data distribution. Therefore, directly applying previous transfer learning methods is inadequate, and some preprocessing of the dataset is necessary. Existing transfer learning methods mainly use over sampling and under sampling strategies \cite{b12, b13, b15, b16} to balance the number of samples for each label when dealing with imbalanced data distributions. The under sampling strategy, which directly removes some data points, may lead to the loss of important information. over sampling may introduce additional noise when the dataset already has significant noise, potentially degrading the training performance. We aim to develop a more nuanced approach that makes more efficient use of the dataset.

\section{METHOD AND APPROACH}
In this section, we first derive the expression for the Loss Function with Weight Filter and provide a theoretical analysis of its advantages compared to existing strategies. We then present the specific implementation algorithm for the Learning Weight Filter.

\begin{figure*}[ht]
    \centering
    \includegraphics[width=0.85\textwidth]{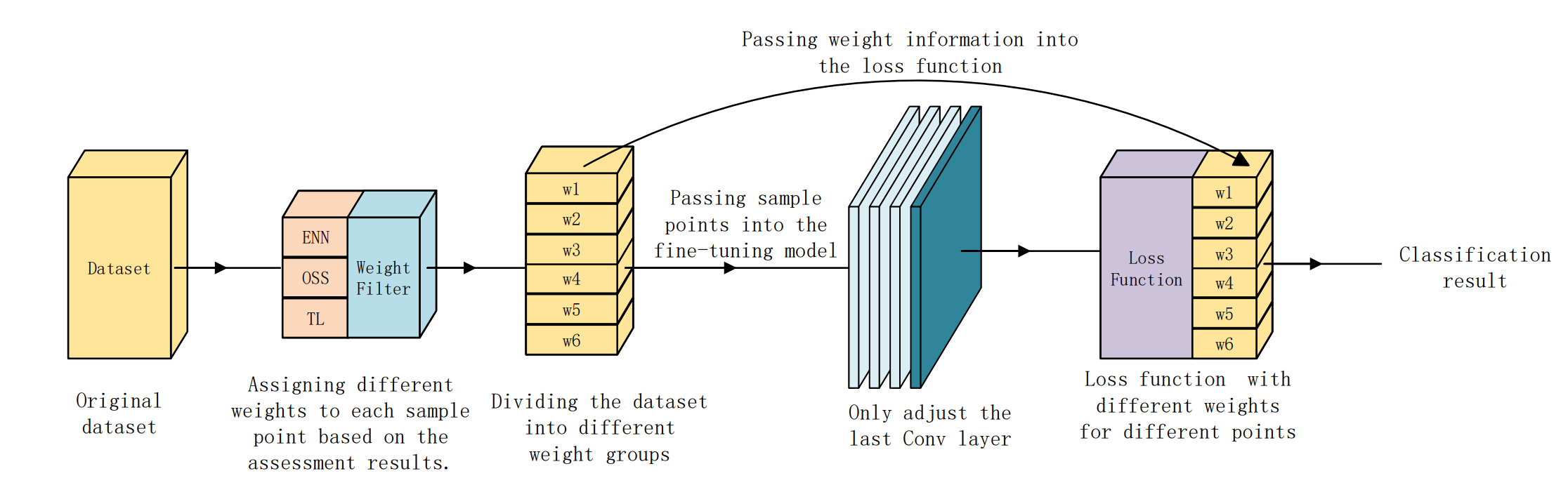}
    \caption{The fine-tuning process using loss function with weight filter(ResNet-34-based)}
    \label{fig:pipeline}
\end{figure*}

\subsection{Mathematical Formulation}
Consider a loss function of the form
\begin{equation}
L(w) = \frac{1}{N} \sum_{i=1}^{N} \ell(y_i, f(x_i; w)),
\end{equation}

\noindent where $N$ represents the total number of samples, $x_i$ is the feature vector of the $i$-th sample, $y_i$ is the true label of the $i$-th sample, $\omega$ denotes the model parameters, $f(\cdot)$ represents the predicted label for a sample by the model, and $\mathcal{L}(\cdot)$ denotes the loss function for an individual sample. The expression for the gradient update is:

\begin{equation}
w \leftarrow w - \eta \nabla_w L(w),
\end{equation}

\begin{equation}
\nabla_w L(w) = \frac{1}{N} \sum_{i=1}^{N} \nabla_w \ell(y_i, f(x_i; w)),
\end{equation}

\noindent where $\eta$ represents the learning rate.
This type of loss function is commonly used in deep learning models where the dataset is abundant and the sample distribution is balanced. However, when the noise in the samples is too high, the loss function values for noisy samples can greatly exceed those of valuable samples. This causes the model to spend excessive effort fitting these less important samples during gradient updates, resulting in poor learning performance. Additionally, when the sample distribution is imbalanced, the value of the loss function is primarily determined by the majority class samples, leading to poor learning outcomes for the minority class samples.

Existing solutions to such problems primarily involve over sampling and under sampling, where a set of artificially engineered sample points are incorporated or some potentially noisy samples are removed to filter out noise and balance the number of different types of samples. Taking under sampling as an example, its loss function is as follows:
\begin{equation}
L(w) = \frac{1}{N_{\text{keep}}} \sum_{i \in S_{\text{keep}}} \ell(y_i, f(x_i; w)),
\end{equation}

\noindent where $N_{\text{keep}}$ represents the number of samples after under sampling, and $S_{\text{keep}}$ denotes the set of samples retained after under sampling.

This method effectively addresses the issues of imbalanced sample distribution and excessive noise. However, by completely removing certain samples, it prevents those samples from contributing to the neural network's learning process. This may result in the loss of valuable information contained in the dataset, making it a less optimal solution.

In response, we propose a filtering algorithm that categorizes the samples passing through the filter into different weight classes based on their importance. For each weight class, a different weight is assigned to the loss function, allowing it to focus on learning key samples while still considering noisy samples. We refer to this approach as the Loss Function with Weight Filter. Suppose we divide the samples into $k$ different weight classes, then the loss function can be expressed as:

\begin{equation}
L(w) = \frac{1}{N} \sum_{j=1}^{k} \alpha_j \sum_{i \in S_j} \ell(y_i, f(x_i; w)),
\end{equation}

\noindent where $S_j$ represents the set of samples in the $j$-th weight class, $\alpha_j$ denotes the weight value for that class, and the different sample sets $S_j$ are mutually exclusive. The union of these sets forms the entire sample set $S$.
\begin{equation}
S = \bigcup_{j=1}^{k} S_j \quad \text{and} \quad S_i \cap S_j = \emptyset \quad \text{for} \quad i \neq j.
\end{equation}

Thus, the gradient update expression for this loss function is:
\begin{equation}
w \leftarrow w - \eta \cdot \frac{1}{N} \sum_{j=1}^{k} \alpha_j \sum_{i \in S_j} \nabla_w \ell(y_i, f(x_i; w)),
\end{equation}

\begin{equation}
\nabla_w L(w) = \frac{1}{N} \sum_{j=1}^{k} \alpha_j \sum_{i \in S_j} \nabla_w \ell(y_i, f(x_i; w)).
\end{equation}

By constructing this new loss function strategy, the training strategy during fine-tuning is adjusted as follows: First, the dataset is processed through the Weight Filter to obtain weights. Then, the dataset is fed into the model that needs fine-tuning. Next, we regard the original loss function and the weights collectively as a new loss function. The results from this new loss function are then used for backpropagation. Finally, the fine-tuned model is obtained. The overall process is illustrated in Fig.\ref{fig:pipeline}.

\subsection{Weight Assignment Algorithm}
The algorithm of the weight filter consists of the following steps: First, the algorithm applies different types of under sampling methods to the sample set, generating retained and removed sample sets. Then, it calculates the occurrence frequency of each sample across the different under sampling results and assigns corresponding weight values to the samples based on a predefined set of weights. Finally, the algorithm returns the weight vector, where the weight of each sample is adjusted according to its importance in the under sampled sets.The specific algorithm workflow is as follows:

\begin{algorithm}
\caption{Weight Assignment Algorithm: Dividing the dataset into different weight classes based on the importance of the samples.}
\label{alg:filter}
\begin{algorithmic}[1]
\REQUIRE The sample set $X = [x_1, x_2, \dots, x_n]$ consisting of $n$ sample points.
\REQUIRE The weight values from different weight sets, denoted as $\alpha = [\alpha_1, \alpha_2, \dots, \alpha_m]$, where $0 \leq \alpha_1 \leq \alpha_2 \leq \dots \leq \alpha_m \leq 1$.
\REQUIRE The set of algorithms $\mathcal{F} = [F_1 (\cdot), F_2 (\cdot), \dots, F_k (\cdot)]$ consisting of under sampling algorithmswith different categories or different initial parameter values.
\ENSURE The weight vector $\mathbf{\omega} = [\omega_1, \omega_2, \dots, \omega_n]$ consisting of the weights of each sample point.

\FOR{$F_k(\cdot)$ in $\mathcal{F}$}
    \STATE $X_{k\text{;}keep} = F_k(X)$
\ENDFOR

\FOR{$x_i$ in $X$}
    \STATE $number = \sum_{j=1}^m (x_i \in X_{k\text{;}keep})$
    \STATE $\omega_i = \alpha_{number}$
\ENDFOR

\RETURN $\omega = [\omega_1, \omega_2, \dots, \omega_n]$

\end{algorithmic}
\end{algorithm}

\section{EXPERIMENT AND RESULT}
This section is divided into three parts. First, we provide an overview of the two datasets used in the experiments. Next, we analyze the feasibility of using a transfer learning algorithm from a theoretical perspective by calculating several statistical features of the two datasets. Finally, we address the problem using the strategy proposed in this paper.

\begin{table*}[h!]
\centering
\begin{tabular}{l|cccc}
\textbf{Accuracy/F1 Score} & LabelSmooth & Focal\cite{b18} & FocalWithLogistics\cite{b18} & LabelSmoothFocal\cite{b19} \\
\hline
No Sampling & 62.73\% / 50.83\% & 64.85\% / 48.45\% & 60.06\% / 57.73\% & 60.98\% / 52.94\% \\
\hline
Over Sampling(Random) & 81.45\% / 72.34\% & 82.67\% / 70.89\% & 79.98\% / 71.45\% & 83.12\% / 73.22\% \\
Over Sampling(Smote\cite{b12}) & 80.67\% / 70.45\% & 83.78\% / 72.56\% & 79.45\% / 73.11\% & 82.34\% / 71.23\% \\
Over Sampling(Adasyn\cite{b13}) & 81.12\% / 73.67\% & 82.98\% / 70.12\% & 80.23\% / 72.89\% & 79.78\% / 69.45\% \\
\hline
Under Sampling(Random) & 83.56\% / 74.32\% & 82.15\% / 71.29\% & 84.92\% / 73.11\% & 83.45\% / 75.67\% \\
Under Sampling(TL\cite{b14}) & 81.78\% / 72.56\% & 84.63\% / 76.34\% & 86.23\% / 78.12\% & 80.98\% / 74.34\% \\
Under Sampling(ENN\cite{b15}) & 85.79\% / 79.18\% & 82.47\% / 76.89\% & 87.56\% / 77.56\% & 85.34\% / 76.45\% \\
Under Sampling(OSS\cite{b16}) & 80.24\% / 73.45\% & 84.78\% / 78.23\% & 85.67\% / 79.12\% & 83.24\% / 75.89\% \\
\hline
\textbf{FilterLoss(OSS)} & 89.34\% / 80.45\% & 87.56\% / 79.23\% & 91.23\% / 82.67\% & 88.78\% / 81.12\% \\
\textbf{FilterLoss(ENN)} & 90.12\% / 78.89\% & 86.78\% / 83.45\% & \textbf{\textcolor{red}{92.34\% / 84.12\%}} & 87.89\% / 79.78\% \\
\textbf{FilterLoss(ENN \& OSS)} & 91.45\% / 80.67\% & 89.78\% / 82.12\% & 86.34\% / 79.89\% & 88.45\% / 81.56\% \\
\end{tabular}
\caption{Performance comparison of different sampling strategies \cite{b12, b13, b14, b15, b16, b17, b18, b19}}
\label{result}
\end{table*}

\subsection{Dataset}
We utilized two datasets in the experiments. The SHL-2018 dataset\cite{b11} contains 16,310 training samples and 5,698 testing samples. Each sample has 12 dimensions. The vector for each dimension consists of 6,000 floating-point numbers, obtained through sampling at a frequency of 100Hz over one minute. The dataset is categorized into eight labels: still, walking, running, bicycling, driving, train, subway, and bus.The Beijing 2024 dataset\cite{b10} sampled a total of 3,259 data points, maintaining the same sample format as the 2018 dataset. It includes six categories of labels, which are still, walking, running, bicycling, subway, and bus, aligning with those categories shared with the 2018 dataset. This approach ensures comparability and maintains the integrity of the analysis by preventing data leakage between training and testing sets during the experiments. To validate the generalizability across the two datasets, samples were uniformly processed to only include the six overlapping label categories from both datasets.

In all subsequent experiments, the baseline model is ResNet\cite{b10}, and the default loss function is cross-entropy\cite{b10}. Other variables are also kept constant.

\begin{figure}[h!]
\centerline{\includegraphics[width = 5cm]{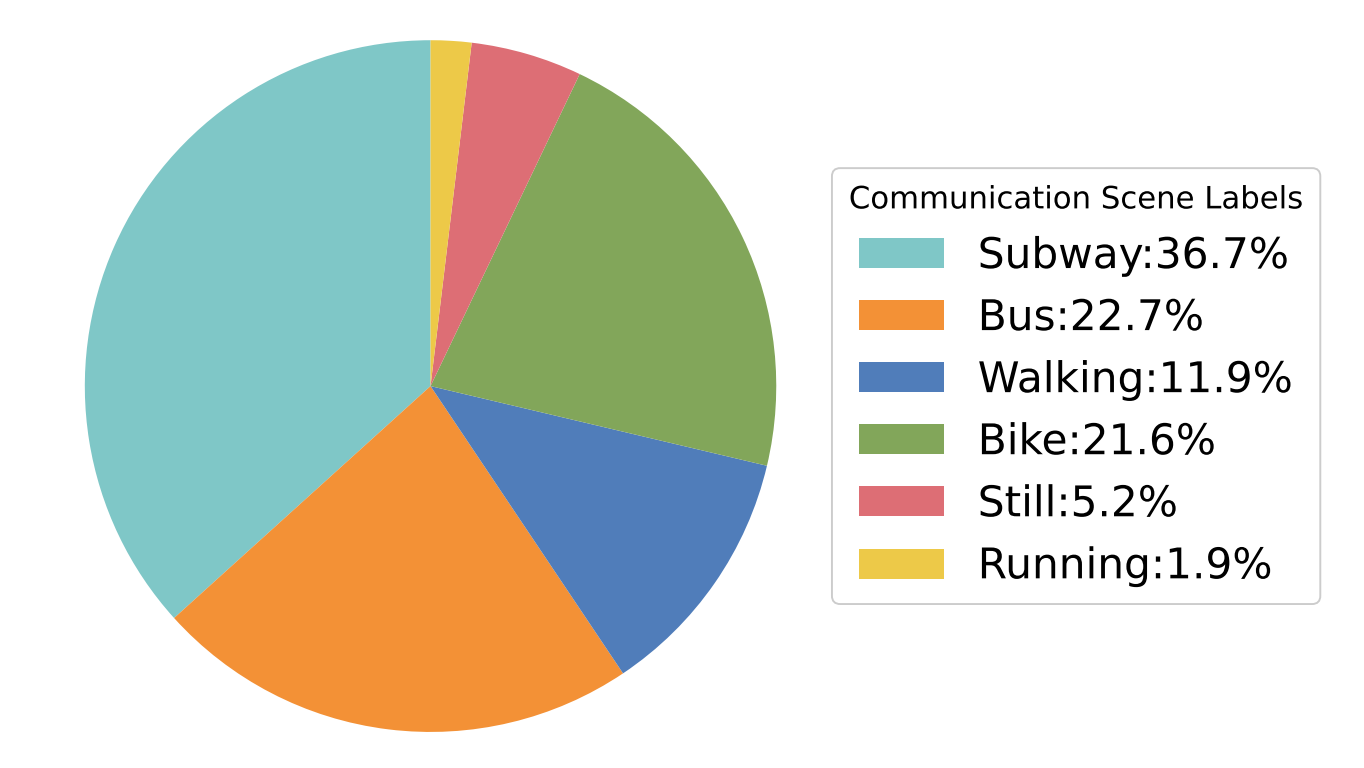}}
\caption{The proportion of data for different labels in Beijing-2024 dataset(highly imbalanced)}
\label{pie}
\end{figure}

\subsection{Data Characteristics of Communication Scenes and Noises}

First, we illustrate in Fig. \ref{pie} that the Beijing-2024 dataset exhibits a significant data imbalance issue, which is not uncommon in the collection of datasets for communication scene recognition.

\begin{figure}[htbp]
    \centerline{\includegraphics[width = 7cm]{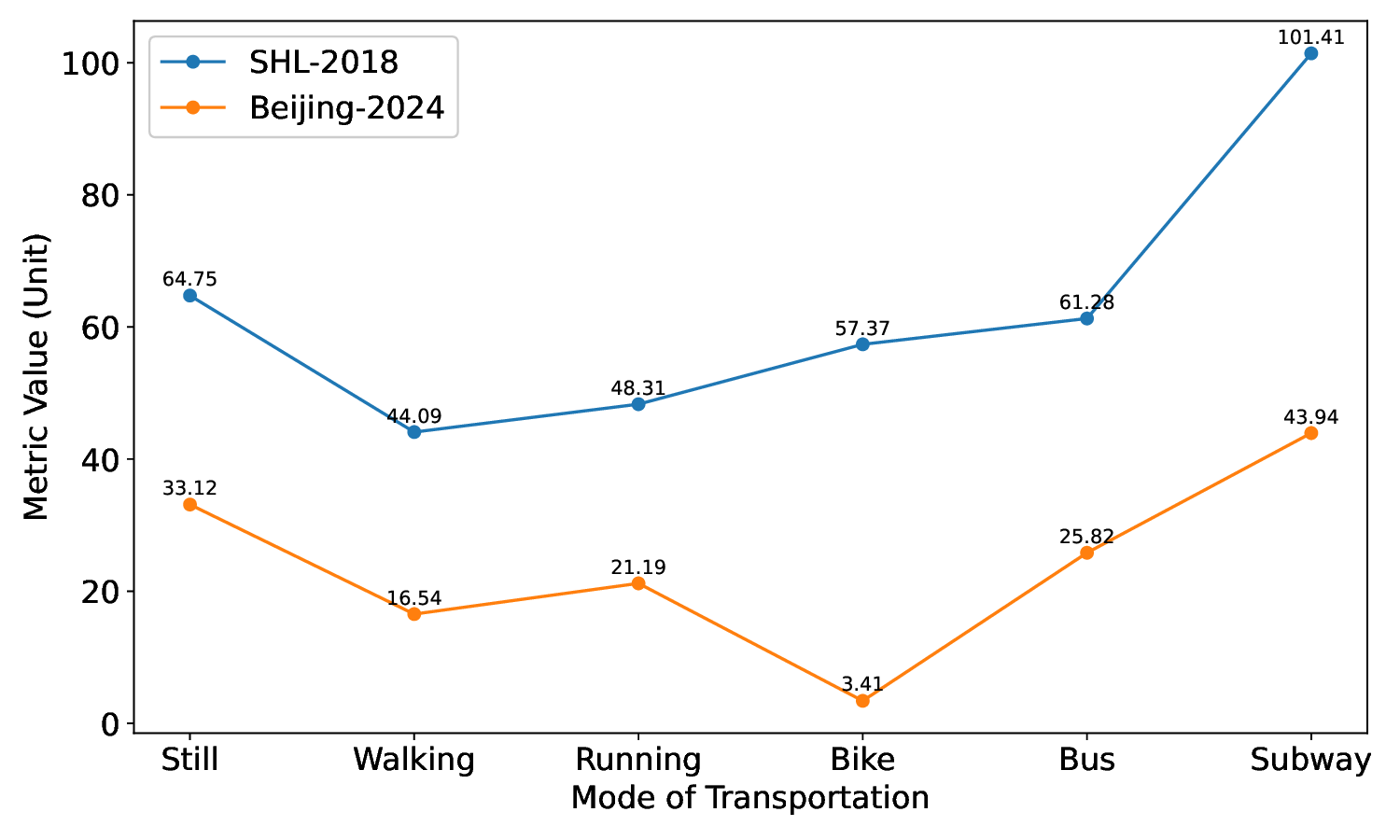}}
    \caption{ Euclidean distance of two datasets under the same scene labels}
    \label{Eu}
\end{figure}

\begin{figure}[htbp]
    \centering
    \centerline{\includegraphics[width = 7cm]{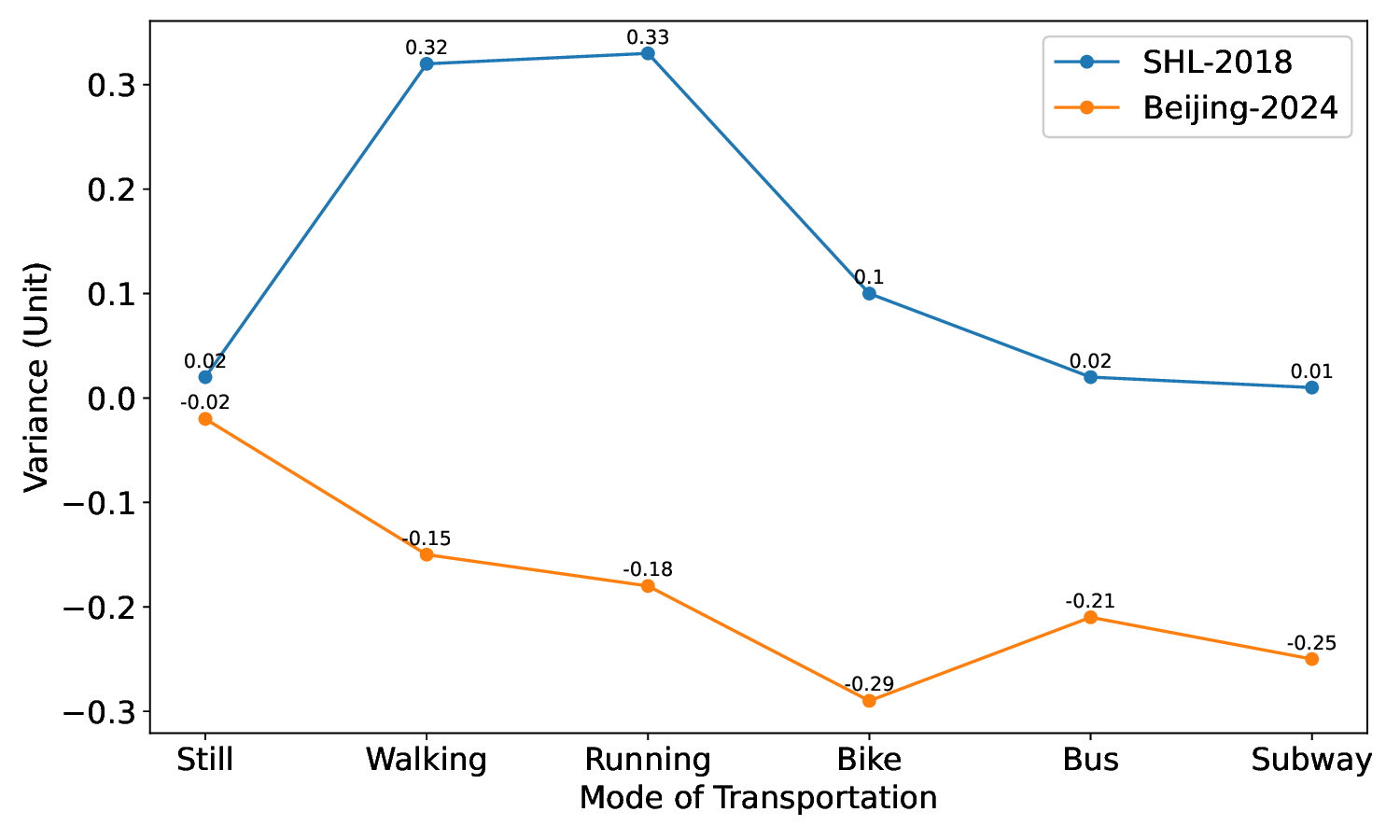}}
    \caption{Cosine similarity of two datasets under the same scene labels}
    \label{Cos}
\end{figure}

Subsequently, we compared the clustering of data points under the same scene labels in the two datasets by quantifying the similarity between data points using two different metrics: Euclidean distance and cosine similarity. Results are presented in Fig. \ref{Eu} and Fig. \ref{Cos}.

The overall Euclidean distance in the Beijing-2024 dataset is lower than in the SHL2018 dataset, but the degree of reduction varies across labels. The cosine similarity in the Beijing-2024 dataset is negative, indicating a smaller drift relative to the origin; whereas, in the SHL-2018 dataset, the cosine similarity is positive, indicating a larger drift relative to the origin. This indicates a higher background noise in the SHL-2018 dataset. The distance and similarity figures show that the average noise of the datasets exhibits a time-varying decreasing trend as well. However, it's also clear that the noise reduction is generally consistent across many scene labels. This suggests that while noise patterns vary, the signal patterns between the two datasets are relatively stable. Therefore, applying transfer learning is indeed possible in addressing the generalization question.

We have demonstrated that it is feasible to train a model using 2018 open-source data and then migrate it to a specific application scene using fine-tuning techniques. Next, we will try out our new fine-tuning strategy through experiments. Our goal is to achieve good accuracy within a short tuning time, utilizing minimal and easily collected data.

\subsection{Fine-tuning with FilterLoss}
In this experiment, for the model pre-trained on the SHL-2018 dataset, we adopted a fine-tuning strategy where all preceding convolutional layers are frozen, and only the last convolutional layer is adjusted. Additionally, we applied the FilterLoss strategy mentioned in this paper.

We employed various over sampling and under sampling strategies, as well as the proposed FilterLoss strategy, for training. We also tested different loss functions, with each model being trained for 10 epochs. The restored accuracy and F1 score are shown in Table \ref{result}(The parentheses following FilterLoss Sampling represent the quality assessment criteria of sample points by the Weight Assignment Algorithm). The baseline model, which is the original model before applying transfer learning, has an accuracy of 83.26\% and a F1 score of 79.82\%. The red marks indicate the category with the highest accuracy and F1 score across all models, while the blue marks represent the highest accuracy and F1 score within the same strategy. The analysis shows that using over sampling and under sampling strategies performs better than directly using the dataset, with under sampling being more effective than over sampling. The proposed FilterLoss strategy outperforms both over sampling and under sampling strategies. This indicates that our method effectively addresses the problem at hand.

The performance improvement brought by using FilterLoss is due to its ability to focus attention on more valuable sample points, similar to existing sampling strategies, while still maintaining appropriate attention to less important, noisier samples. Unlike under sampling, it does not discard these samples outright, and unlike over sampling, it does not generate non-existent samples, which could introduce more noise.

\begin{figure}[htbp]
    \centering
    \centerline{\includegraphics[width = 9cm]{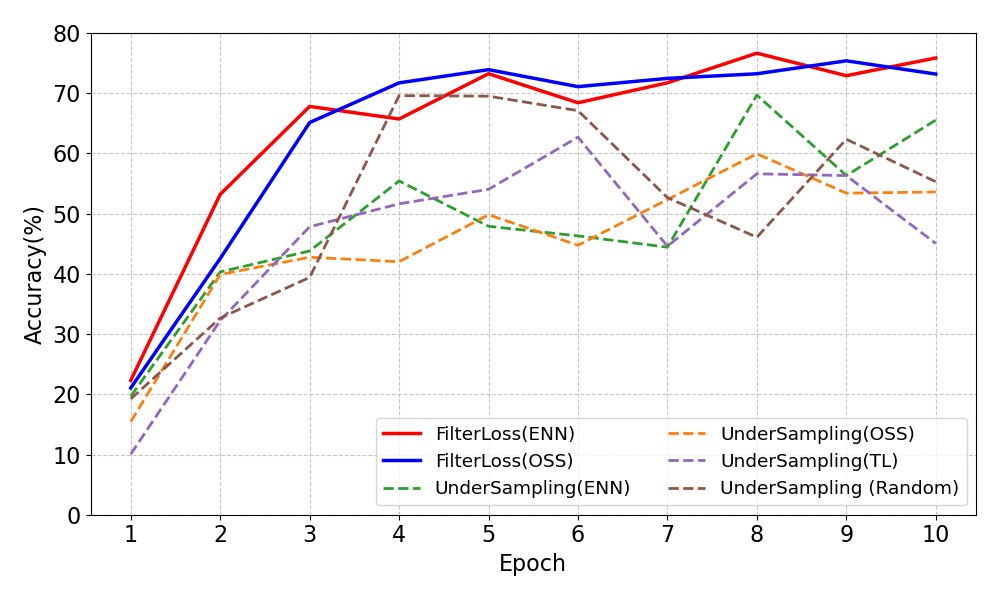}}
    \caption{Accuracy of two strategies: FilterLoss outperforms other strategies}
    \label{accuracy}
\end{figure}

We also found that our FilterLoss strategy demonstrated greater stability in the experiments. We plotted the accuracy of various models over the first 10 epochs using both the FilterLoss and Under Sampling strategies (Fig. \ref{accuracy}). The results showed that models using the FilterLoss strategy were relatively more stable.This increased stability is due to the FilterLoss strategy not discarding certain sample points directly as Under Sampling does. The samples discarded by the Under Sampling strategy are often located near class boundaries, leading to improved handling of boundary-related issues by the FilterLoss strategy. Moreover, the lower weight values assigned to boundary points prevent the model from allocating excessive attention to these areas.

\section{CONCLUSION}
We propose FilterLoss, a "soft" sampling strategy combined with loss function that assigns different weight values to the loss function for each sample point. By assigning different weight coefficients to the loss function of each sample point \cite{b66}, the model can focus primarily on the more informative samples, while still maintaining appropriate attention to noisier data points. We also developed a corresponding weight assignment algorithm to determine the appropriate weight for each sample point. We propose this strategy to address the issues of insufficient data and imbalanced data distribution in communication scene recognition.

\begin{table}[h!]
\centering
\begin{tabular}{c|c|c}
\toprule
Method & Accuracy & F1 Score\\
\midrule
No Sampling ResNet & 60.06\% & 57.73\% \\
Over Sampling ResNet & 80.23\% & 72.89\% \\
Under Sampling ResNet & 84.92\% & 73.11\% \\
\textbf{FilterLoss ResNet} & \textbf{92.34\%} & \textbf{84.12\%} \\
\bottomrule
\end{tabular}
\caption{Comparison of different methods}
\label{result}
\end{table}

By using FilterLoss, we were able to restore the accuracy of transfer learning to the original benchmark of 92.34\%, while significantly reducing the labor and computational costs. Data collection can be performed using a smartphone, a tablet or a smart watch, enabling personalized models for each device. Compared to directly using imbalanced datasets or datasets processed by over sampling and under sampling, the model trained with our proposed method achieves relatively higher accuracy and more stable results.

Most mobile communication devices meet the computational requirements of our method, enabling its deployment across various mobile terminals. The use of transfer learning allows it to adapt to a wide range of mobile devices and sensors, providing users with personalized services.

In this paper, the weights were obtained through cross-validation to achieve optimal values, but this approach limits the assignment of weights to only a few broad categories of samples. By learning the weights of the loss function as parameters, we could assign the most suitable weight to each individual sample point, thereby improving performance.

\bibliographystyle{IEEEtran}
\bibliography{ref}

\end{document}